
\magnification=\magstep1
\hsize=13cm
\vsize=20cm
\overfullrule 0pt
\baselineskip=13pt plus1pt minus1pt
\lineskip=3.5pt plus1pt minus1pt
\lineskiplimit=3.5pt
\parskip=4pt plus1pt minus4pt

\def\negenspace{\kern-1.1em}


\newcount\secno
\secno=0
\newcount\susecno
\newcount\fmno\def\z{\global\advance\fmno by 1 \the\secno.
                       \the\susecno.\the\fmno}
\def\section#1{\global\advance\secno by 1
                \susecno=0 \fmno=0
                \centerline{\bf \the\secno. #1}\par}
\def\subsection#1{\medbreak\global\advance\susecno by 1
                  \fmno=0
       \noindent{\the\secno.\the\susecno. {\it #1}}\noindent}
%
\def\sqr#1#2{{\vcenter{\hrule height.#2pt\hbox{\vrule width.#2pt
height#1pt \kern#1pt \vrule width.#2pt}\hrule height.#2pt}}}
\def\square{\mathchoice\sqr64\sqr64\sqr{4.2}3\sqr{3.0}3}
\setbox1=\hbox{$\square$}
\def\dalem{\raise 1pt\copy1}
%


\newcount\refno
\refno=1
\def\y{\the\refno}
\def\myfoot#1{\footnote{$^{(\y)}$}{#1}
                 \advance\refno by 1}


\def\newref{\vskip 1pc 
            \hangindent=2pc
            \hangafter=1
            \noindent}

\def\neq{\hbox{$\,$=\kern-6.5pt /$\,$}}






\newcount\secno
\secno=0
\newcount\fmno\def\z{\global\advance\fmno by 1 \the\secno.
                       \the\fmno}
\def\sectio#1{\medbreak\global\advance\secno by 1
                  \fmno=0
       \noindent{\the\secno. {\it #1}}\noindent}



\null
\bigskip\bigskip\bigskip\bigskip
\centerline{\bf SPIN-DRIVEN INFLATION}
\bigskip  
\bigskip
\bigskip  
\centerline{Yuri N.\ Obukhov$^{*\diamond}$}
\bigskip  
\bigskip  
\centerline{\it Institute for Theoretical Physics, University of  
Cologne, D--50923 K\"oln}  
\centerline{\it Germany}

\bigskip\bigskip  
\bigskip  
\centerline{\bf Abstract} 

Following recent studies of Ford, we suggest -- in the framework of general
relativity -- an inflationary cosmological model with the self-interacting 
spinning matter. A generalization of the standard fluid model is discussed 
and estimates of the physical parameters of the evolution are given. 

\bigskip\bigskip  
\bigskip  
\bigskip\bigskip  

\bigskip\bigskip  
\bigskip  
\bigskip\bigskip  

\vfill

\noindent $^{*})$ Permanent address: Department of Theoretical Physics, 
Moscow State University, 117234 Moscow, Russia.

\noindent $^{\diamond})$ Alexander von Humboldt Fellow.

\eject  

In the recent paper of Ford [1] a new inflationary model was proposed in which
the inflation mechanism was related to the existence of a self-interacting 
vector field, replacing the usual scalars. Such a structure could be explained
by the nonlinear Lagrangian, {\it effectively} arising from a more 
fundamental interactions between the elements of cosmological matter. However,
the vector field of a non-gauge nature does not appear to be a physically 
important object, in particular, not in cosmology. In this paper we use the 
main idea of Ford and propose an alternative mechanism for inflation, which 
is related to the spin of matter. Spin-spin interactions play an important 
role in many field-theoretical models, in particular in torsion theories [2]. 
We will not specify the form of a possible fundamental spin-spin interaction 
(although taking the Poincar\'e gauge approach [3] as the basic model), but 
instead proceed within a similar {\it effective} approach in the framework
of the spinning fluid variational theory. The crucial point of the proposed
generalization is the introduction of a non-linear spin term in the fluid
Lagrangian. It represents the contribution of the effectively averaged 
Poincar\'e rotational gauge fields, e.g., in the sense of integrating away 
the torsion [4].

Let us postulate the Lagrangian for the spinning fluid with self-interaction 
$$
L_{m}=\epsilon (\rho, s, \mu^{ij}) - 
{1\over 2}\rho\mu^{ij}b^{\mu}_{i}(\nabla_{\alpha}b^{\nu}_{j})u^{\alpha}
g_{\mu\nu} 
$$
$$
+ \rho u^{\mu}\partial_{\mu}\lambda_{1} + 
\lambda_{2}u^{\mu}\partial_{\mu}X + \lambda_{3}u^{\mu}\partial_{\mu}s +
\lambda^{ab}(g_{\mu\nu}b^{\mu}_{a}b^{\nu}_{b}- \eta_{ab})+ V(\xi).\eqno(1)
$$
Here the terms with the Lagrange multipliers $\lambda$ describe the usual
constraints [5], imposed on the fluid variables $\rho$ (particle density),
$s$ (specific entropy), $X$ (identity Lin coordinate), $\mu^{ij}$ (specific
spin density), and $b^{\mu}_{a}$ (material tetrad, with $b^{\mu}_{0}=u^{\mu}$ 
the 4-velocity of elements). [Indices from the middle of the Latin alphabet 
$i,j=1,2,3$ refer to the local 3-frame defined by the spacelike vectors of 
the material tetrad; the Greek indices $\alpha,\beta,...=0,1,2,3$ refer to
space-time coordinates]. The variable
$$
\xi={1\over 2}\rho^2\mu^{ij}\mu_{ij}\eqno(2)
$$
is the square of the spin density of the matter. The function $V(\xi)$ can 
be thought as a kind of the effective potential which arises due to the 
underlying fundamental interactions between particles with spin. A particular 
example is given by the Einstein-Cartan theory [2] in which $V$ is a linear 
function of $\xi$.

The equations of motion of the fluid and the gravitational Einstein equations
are derived from the variation of (1) with respect to the fluid and 
gravitational variables. This yields the following modified energy-momentum 
tensor of the spinning fluid
$$
T_{\mu\nu} = \epsilon u_{\mu}u_{\nu} - p h_{\mu\nu} 
- 2(g^{\alpha\beta}+u^{\alpha}u^{\beta}) \nabla _{\alpha}
(u_{(\mu}S_{\nu)\beta}) +
$$
$$
+ g_{\mu\nu}V-h_{\mu\nu}2\xi V^{\prime},\eqno(3)
$$
where $(^{\prime})$ denotes the derivative with respect to $\xi$, 
and, as usually, $p$ is the pressure, $S^{\mu\nu}=-{1\over 2}\rho b^{\mu}_{i}
b^{\nu}_{j}\mu^{ij}$ is the spin density, and the projector on the subspace 
orthogonal to 4-velocity is $h^{\mu}_{\nu}=\delta^{\mu}_{\nu} - 
u^{\mu}u_{\nu}$.

The dynamics of the spin is not changed, and the standard equation of motion is
valid (which folllows from the variation of (1) with respect to the material
tetrad):
$$
\nabla _{\mu}(u^{\mu}S_{\alpha\beta}) = 
u_{\alpha}u^{\lambda}\nabla _{\mu}(u^{\mu}S_{\lambda\beta}) -
u_{\beta}u^{\lambda}\nabla _{\mu}(u^{\mu}S_{\lambda\alpha}).\eqno(4)
$$
For concretness, in (1)-(4) we consider the case of the ordinary spinning 
fluid. Eq.(4) shows that the motion of spin is not affected by the 
nonlinearity: this is essentially a rotation, precession. The same, however, 
is true also for the generalized fluid with magnetic moment and electric 
charge in the electromagnetic field. 

But the dynamics of the fluid itself is changed. It is described by the 
conservation of the energy-momentum, and for (3) this reads
$$
\nabla_{\mu} T^{\mu}_{\ \nu} =
u_{\nu}(u^{\mu}\nabla_{\mu}\epsilon + \epsilon\nabla_{\mu}u^{\mu} + 
p\nabla_{\mu}u^{\mu})
$$
$$
 - h^{\mu}_{\ \nu}\nabla_{\mu}(p + 
2\xi V^{\prime})+(p + \epsilon +2\xi V^{\prime})a_{\nu} 
+\nabla_{\nu}V + u_{\nu}(\nabla_{\mu}u^{\mu})2\xi V^{\prime} 
$$
$$
+ 2 u^{\mu}S_{\nu\lambda}\nabla_{\mu}a^{\lambda} + 
R_{\alpha\beta\mu\nu}u^{\mu}S^{\alpha\beta} = 0.\eqno(5)
$$
Standard projections on the $u^{\mu}$ and the orthogonal directions yield
$$
(p + \epsilon)\nabla_{\mu}u^{\mu} + u^{\mu}\nabla_{\mu}\epsilon = 0,\eqno(6)
$$
$$
(p + \epsilon + 2\xi V^{\prime})a_{\nu} - h^{\mu}_{\ \nu}\nabla_{\mu}(p +
2\xi V^{\prime} - V) +
2 S_{\nu\mu}u^{\lambda}\nabla_{\lambda}a^{\mu} + 
R_{\alpha\beta\mu\nu}u^{\mu}S^{\alpha\beta} = 0.\eqno(7)
$$

Now we are in a position to proceed in a way typical for the inflational
approach: it is necessary to specify the ``effective potential" so that the
inflationary stage becomes possible. 

It is straightforward to analyse the modification of the standard cosmologies.
Let us consider the special case of a flat model with the line element
$$
ds^2=dt^2 -R^2 (t)(dx^2 + dy^2 + dz^2).
$$
With the self-interacting spinning fluid (3), the Einstein equations become
$$
3{\dot{R}^2 \over R^2}=\kappa (\epsilon + V),\eqno(8)
$$
$$
-(2{\ddot{R} \over R} + {\dot{R}^2 \over R^2})=\kappa (p-V+2\xi V^{\prime}).
\eqno(9)
$$
Instead of (9) it is more convenient to use the conservation law (6), which
reads as usual:
$$
\dot{\epsilon}+3{\dot{R} \over R}(\epsilon + p)=0.\eqno(10)
$$
Clearly, eqs. (8) and (10) must be supplemented by the equation of state
$p=p(\epsilon)$. Notice that eq. (9) follows from (8) and (10).

It is important to note the specific behaviour of the argument $\xi$ of the 
potential. It is the square of the spin ($\xi=2S_{\mu\nu}S^{\mu\nu}$), and 
from the equations of motion (4) one finds
$$
\xi (R(t)) = {const \over R^6}.\eqno(11)
$$
Thus, during the expansion of the universe,
$$
V(\xi)\rightarrow V(0).
$$
Following the lines of reasoning of Ford (who studied inflation with a vector 
field), one can discuss now the possible form of the potential $V$. Most 
appropriate seems to be what is called in [1] the ``chaotic inflation" model, 
in which inflation occurs for large $\xi$ and reheating as $\xi\rightarrow 0$.
A potential which provides such a possibility reads
$$
V(\xi) = V_{0}(1-e^{-\alpha\xi}),\eqno(12)
$$
where the parameters $V_{0},\alpha$ are chosen from the estimates for the
characteristics of the inflationary period.

It is clear from (8) that the cosmological evolution --- the dynamics of
the scale factor $R(t)$ --- depends on the relative value of the $\epsilon$
and $V$. Namely, when $V$ is much smaller than the energy density, the
evolution proceeds as in the standard cosmology. For (12) this occurs 
inevitably as $\xi$ tends to zero. However, if $\epsilon$ is less than $V$
and the latter is approximately constant, the inflation occurs. Let us for
definiteness choose the equation of state $p={\epsilon \over 3}$. Then from
(10) 
$$
\epsilon = {const \over R^4}.\eqno(13)
$$
Clearly, at the initial stages (when $R\sim 0$) the energy density is 
greater than the constant $V_{0}$ (which is the value of the spin potential
at early times, when $\xi$ is very large). We can roughly consider the
moment when $\epsilon$ becomes equal $V_{0}$ to be the starting point of
inflation. Let us denote the values of the energy and spin density at that
moment as 
$$
\epsilon_{0},\ \ \ \ \ S_{0}.
$$
As the end of the inflationary stage (approximately) could be considered the 
point $t=\tau$ at which the (decreasing) energy density $\epsilon(\tau)$ again
becomes equal to some $V(\xi(\tau))$ (which is decreasing much more fastly). 
Let us denote the ratio of the scale factors at the beginning and the end of 
the inflationary stage by
$$
E={R_{0} \over R(\tau)}.\eqno(14)
$$
One has $\xi(\tau)=\xi_{0}E^6=S_{0}^{2}E^6,\ \ \epsilon(\tau)=\epsilon_{0}E^4$.
Hence from (12) one can derive the relation between these quantities:
$$
\epsilon(\tau)=V(\tau)\ \ \ \rightarrow\ \ \ E^4=
1-e^{-\alpha S^{2}_{0}E^6}\approx \alpha S^{2}_{0}E^6,\eqno(15)
$$
Assuming, that $E^{-1}\approx 10^{20}$, this gives the estimate
for the constant $\alpha$ and the initial spin density,
$$
\alpha S_{0}^2 \approx 10^{40}.\eqno(16)
$$
This is large enough, so at the beginning of the inflation $V$ is constant,
equal to $V_{0}$. Very crudely let us assume that 
$V_{0}=10^{94} {g \over cm^3}$ ($=\epsilon_{0}$), i.e. inflation starts
directly at Planckian values. Then at the end of inflation
$\epsilon(\tau)=10^{14}{g \over cm^3}$ which is about the quark density. The 
duration of the inflational stage is then of order of a Planckian time, 
estimated by
$$
\tau = {20\log 10 \over \sqrt{{\kappa V_{0} \over 3}}}
\approx 6\times 10^{-43}s.\eqno(17)
$$

Thus in this scheme, analogously to the developments of Ford one can
predict an inflation stage during the universe's evolution, which is caused
by a non-linear effective potential arising from a fundamental interaction
of the cosmological matter. In our opinion, the spin non-linearity is much
more natural than the vector field one, originally used in [1].

Let us point out that one can also develop the original idea of Ford into 
a kind of general scheme: it is clear, that the crucial point is to have 
some self-interacting (i.e. non-linear) system coupled to the Einstein 
gravity, then for special choices of a ``potential" (non-linearity) one can 
discover inflation. 

The estimates derived above can be changed and improved for the potentials
different from (12). An interesting problem is to derive the form of the
spin nonlinearity $V$ when the fluid is considered as a semiclassical 
description of a realistic quantum matter. 
\bigskip
\bigskip
{\bf Acknowledgements}
\bigskip
\bigskip
I would like to thank Prof. Friedrich W. Hehl for the careful reading the 
manuscript and useful advice. This research was supported by the Alexander 
von Humboldt Foundation (Bonn).
\vfill\eject
\bigskip
\bigskip
{\bf References}
\bigskip
\newref
[1] L.H. Ford, {\sl Phys. Rev.} {\bf D40} (1989) 967-972. 
\newref
[2] F.W. Hehl, P. von der Heyde, G.D. Kerlick, and J.M. Nester, {\sl Revs.
Mod. Phys.} {\bf 48} (1976) 393; A. Trautman, {\sl Symposia Mathematica} 
{\bf 12} (1973) 139.
\newref
[3] F.W. Hehl, J. Nitsch, and P. von der Heyde, in: {\it General Relativity and
Gravitation: One Hundred Years after the Birth of Albert Einstein}, Ed. 
A.Held (Plenum: New York, 1980) vol. 1, 329; 
K.A. Pilch, {\it Lett. Math. Phys.}, {\bf 4} (1980) 49;
A.A. Tseytlin, {\it Phys. Rev.}, {\bf D26} (1982) 3327;
J. Hennig, and J. Nitsch, {\it Gen. Rel. Grav.}, {\bf 13} (1981) 947;
L.K. Norris, R.O. Fulp, and W.R. Davis, {\it Phys. Lett.}, {\bf A79}
(1980) 278.
\newref
[4] A.A. Tseytlin, {\it J. Phys.} {\bf A15} (1982) L105.
\newref
[5] Yu.N. Obukhov, and V.A. Korotky, {\sl Class. Quantum Grav.} {\bf 4} (1987)
1633; Yu.N. Obukhov, and O.B. Piskareva {\sl Class. Quantum Grav.} {\bf 6} 
(1989) L15.

\end